\documentclass[12pt]{article}
\usepackage{float}
\usepackage{amssymb, amsmath, amsthm}
\usepackage{mathrsfs}
\usepackage{tabulary}
\usepackage{rotating}
\usepackage{setspace}
\usepackage{multirow}
\usepackage{paralist}
\usepackage{listings}
\usepackage[T1]{fontenc}
\usepackage[utf8]{inputenc}
\usepackage[english]{babel}
\usepackage{authblk}
\usepackage{etoolbox}
\usepackage{dsfont}
\usepackage{hyperref}
\usepackage{subcaption}
\usepackage{amsmath}
\usepackage{graphicx}
\usepackage{enumerate}
\usepackage{url}
\newtheorem{definition}{Definition}
{
  \theoremstyle{definition}
  \newtheorem{assumption}{Assumption}{}
}

\usepackage{longtable}
\usepackage{natbib}
\usepackage[usenames,dvipsnames]{color}
\usepackage{bibentry}
\DeclareMathOperator*{\argmin}{arg\,min}
\renewcommand{\P}{\mathsf{P}}
\newcommand{\E}{\mathsf{E}}
\newcommand{\B}{\mathsf{B}}
\renewcommand{\d}{\mathsf{d}}
\newcommand{\dd}{\mathrm{d}}

\newcommand{\p}{\mathsf{p}} \newcommand{\q}{\mathsf{q}}
\newcommand{\g}{\mathsf{g}} \newcommand{\e}{\mathsf{e}}
\newcommand{\uu}{\mathsf{u}} \newcommand{\h}{\mathsf{h}}
\newcommand{\vv}{\mathsf{v}} \newcommand{\rr}{\mathsf{r}}
\newcommand{\bb}{\mathsf{b}}
\newcommand{\one}{\mathds{1}}

\newcommand\BibTeX{{\rmfamily B\kern-.05em \textsc{i\kern-.025em b}\kern-.08em
T\kern-.1667em\lower.7ex\hbox{E}\kern-.125emX}}

\usepackage{caption}

\IfFileExists{upquote.sty}{\usepackage{upquote}}{}
\usepackage{url} 

\newcommand{\blind}{1}

\newcommand{\indep}{\mbox{$\perp\!\!\!\perp$}}

\begin{document}

\def\spacingset#1{\renewcommand{\baselinestretch}%
{#1}\small\normalsize} \spacingset{1}


\if1\blind
{
  \title{\bf When the ends don't justify the means: Learning a treatment strategy to prevent harmful indirect effects        }
  \author{Kara E. Rudolph\thanks{
    The authors gratefully acknowledge \textit{R00DA042127 from the National Institute on Drug Abuse}}\hspace{.2cm}\\
    Department of Epidemiology, Columbia University\\
    and \\
    Iv\'an D\'iaz \\
    Department of Population
  Health Sciences, Weill Cornell Medicine}
  \maketitle
} \fi

\if0\blind
{
  \bigskip
  \bigskip
  \bigskip
  \begin{center}
    {\LARGE\bf When the ends don't justify the means:\\
    \vspace{.2cm}
     Learning a treatment strategy to prevent harmful indirect effects}
\end{center}
  \medskip
} \fi

\bigskip
\begin{abstract}
There is a growing literature on finding so-called optimal treatment rules, which are rules by which to assign treatment to individuals based on an individual's characteristics, such that a desired outcome is maximized.  A related goal entails identifying individuals who are predicted to have a harmful indirect effect (the effect of treatment on an outcome through mediators) even in the presence of an overall beneficial effect of the treatment on the outcome. In some cases, the likelihood of a harmful indirect effect may outweigh a likely beneficial overall effect, and would be reason to caution against treatment for indicated individuals. We build on both the current mediation and optimal treatment rule literature to propose a method of identifying a subgroup for which the treatment effect through the mediator is harmful. Our approach is nonparametric, incorporates post-treatment variables that may confound the mediator-outcome relationship, and does not make restrictions on the distribution of baseline covariates, mediating variables (considered jointly), or outcomes. We apply the proposed approach to identify a subgroup of boys in the Moving to Opportunity housing voucher experiment who are predicted to have harmful indirect effects, though the average total effect is beneficial.    
\end{abstract}

\noindent%
\vfill

\newpage
\spacingset{1.5} 
\section{Introduction}

When individuals differ in their responses to a treatment or intervention, it may be desirable to use individual-level information to predict for whom treatments/interventions may work well versus not. This goal has fueled the growing literature on finding so-called optimal individualized treatment rules/regimes, which assign treatment based on an individual's characteristics such that a desired outcome is maximized \citep{Murphy03}. For a single point in time and a binary treatment, the optimal treatment rule is often defined as the maximizer over all possible treatment rules of the counterfactual mean outcome observed under a hypothetical implementation of the rule \citep{Murphy03,Robins03a}.

\citet{nabi2018estimation} and \citet{shpitser2018identification} recently extended these methods to 
    finding a treatment rule that optimizes a path-specific effect. \citet{nabi2018estimation} focus, in particular, on optimizing a direct effect (an effect of treatment on an outcome, not through a mediator); 
    in their application, the direct effect of interest is the effect of a treatment not operating through adherence. 
    A related goal entails 
    identifying individuals who are predicted to have a harmful indirect effect (the effect of treatment on an outcome through mediators). Such a goal would be of interest if the process by which an effect of a treatment or intervention might occur may be traumatic or otherwise harmful, or if the harmful indirect effect is indicative of more general harmful social processes, and consequently desirable to avoid. It is possible for harmful indirect effects 
    to be present even if the overall effect of treatment on the outcome is beneficial, sometimes referred to as inconsistent mediation \citep{mackinnon2000equivalence}. In some cases, the likelihood of a harmful indirect effect may outweigh a likely beneficial overall effect and would be reason to caution against treatment for indicated individuals.
    
    The Moving to Opportunity study (MTO) presents an example 
    where it may be of interest to identify those with potentially harmful indirect effects. In MTO, families were randomized to receive a Section 8 housing voucher that would allow them to move out of public housing and into a rental on the private market \citep{kling2007experimental}. Participants were followed for 10-15 years and a broad array of outcomes related to health, education and income were assessed. Among the children, boys in the voucher group were generally found to have negative effects on mental health and risk behaviors (though this was not true for all such outcomes) whereas girls were generally found to have positive effects \citep{clampet2011moving,sanbonmatsu2011moving,schmidt2017adolescence,kessler2014associations,rudolph2018mediation,kling2007experimental,rudolph2020helped}. Much of the negative effects were explained by mediators related to the neighborhood and school environments and instability of the social environment \citep{rudolph2020helped}. However, measures of the school and neighborhood environments appeared to generally improve among those receiving the intervention, indicating that the ``objective'' mediator measures of a school's academic performance, poverty level, etc., may in fact be surrogates for harmful processes like discrimination and increased risk of suspensions/expulsions for some \citep{rudolph2020helped}. In the context of MTO, one could argue that a slightly beneficial long-term overall effect of housing voucher receipt on reduced risk of a particular outcome, e.g., adolescent alcohol use, may be outweighed by harmful indirect effects, as these may indicate the occurrence of harmful social processes more generally. 
    
    Such scenarios motivate identifying those predicted to have harmful indirect effects, and for that subgroup, treatment may be reconsidered. In this sense, such subgroup identification is descriptive rather than prescriptive. Although the approach we propose is mathematically grounded in the optimal individualized treatment rule literature, we avoid using the phrase ``treatment rule'' in describing it, because our subgroup identification is not a treatment recommendation (in part because the overall treatment effect may still be positive). 
    A decision to treat (or not) individuals in this subset would likely be based on a comprehensive assessment involving aspects like 
    the predicted direct effect as well as 
    ethics, costs, and qualitative measures.
    
    In this paper, we build on both the current mediation and optimal individualized treatment rule literature to propose a method of identifying a subgroup who may have an unintended 
    harmful indirect effect, and then estimate how the population path-specific effects would hypothetically change if the identified subgroup were not treated. 
    Our approach is nonparametric, incorporates an ensemble of machine learning algorithms in model fitting, incorporates post-treatment variables that may confound the mediator-outcome relationship, and does not make restrictions on the distribution of baseline covariates, mediating variables (considered jointly), or outcomes. We apply the proposed approach to identify a subgroup of boys in the MTO housing voucher experiment who are predicted to have harmful indirect effects, though the average total effect on adolescent alcohol use is beneficial.

\section{Notation}
Let $O=(W,A,Z, M, Y)$ represent observed data, where $W$ is a vector of baseline variables; $A$ is a binary treatment representing housing voucher receipt; $M$ represents mediating variables related to the school, neighborhood, and social environments; $Y$ represents the outcome of alcohol use in adolescence, and $Z$ represents post-treatment variables that may confound the $M\rightarrow Y$ relationship (in the case of MTO, moving with the housing voucher). Let $O_1, ..., O_n$ represent the sample of $n$ i.i.d. observations of $O$. We use capital letters to denote random variables and lowercase letters to denote realizations of those variables. Let $\P$ represent the distribution of $O$. We assume the data are generated according to the following nonparametric structural equation model \citep{pearl2000causality}:
\begin{multline}
    W=f_W(U_W); A=f_A(W,U_A); Z=f_Z(A,W,U_Z); \\ M=f_M(Z, A, W, U_M); 
    Y=f_Y(M, Z, A, W, U_Y),
\end{multline}
where each $U$ represents unmeasured, exogenous factors and each $f$ represents an unknown, deterministic function. 

Our method to identify subsets of individuals at risk of a harmful indirect effect are defined in terms of counterfactual outcomes under interventions on the treatment and mediator. Specifically, we define a counterfactual outcome under interventions that set the treatment and mediator to $(A,M)=(a,m)$ as $Y_{a,m}=f_Y(m, Z_a, a, W, U_Y)$, where $Z_a=f_Z(a, W, U_Z)$ is the counterfactual variable $Z$ observed under an intervention setting $A=a$. Direct and indirect effects are not generally point-identified in the presence of a variable $Z$ that confounds the mediator-outcome relation and that is affected by treatment \citep{avin2005identifiability}. We overcome this problem using so-called \textit{interventional effects} \citep{vanderweele2014effect,vansteelandt2017interventional,zheng2017longitudinal,lok2016defining, lok2019causal,
  rudolph2017robust,didelez2006direct}, which rely on interventions on that set the mediator to a random draw from its counterfactual distribution conditional on baseline variables. Specifically, we let $M_a=f_M(Z_a, a, W, U_M)$, and use $G_a$ to denote a random draw from the distribution of $M_a$ conditional on $W$.

We define $\P f = \int f(o)\dd \P(o)$ for a given function $f(o)$. We assume $\P$  is continuous with respect to
some dominating measure $\nu$ and let $\p$ denote the corresponding probability
density function. We will also use the following notation: $\bb(a,z,m,w)$ denotes $\E(Y\mid  A=a, Z=z, M=m, W=w)$; $\g(a\mid w)$ denotes $\P(A=a\mid  W=w)$; $\e(a\mid m,w)$ denotes $\P(A=a \mid M= m,W=w)$; $\q(z\mid a,w)$ denotes $\P(Z=z\mid A=a,W=w)$; and $\rr(z\mid a,m,w)$ denotes $\P(Z=z\mid A=a,M=m,W=w)$. 

\section{Identification of the subgroup with a predicted harmful indirect effect }
\subsection{Estimand and identification}
\label{sec:ruleident}
Throughout, we use the example from MTO in which a reduction in risk of subsequent adolescent alcohol use is the outcome of interest; thus, negative risk differences are considered ``beneficial'' and positive risk differences are considered ``harmful''. The definition of ``harmful'' for what follows should be changed to reflect the particular research question. 

As noted above, 
 we use population interventional direct and indirect effects. We 
 only provide their definition here, as they have have been described extensively elsewhere \citep{petersen2006estimation,van2008direct,zheng2012targeted,vanderweele2014effect,rudolph2017robust,vanderweele2017mediation,diaz2019non}. 
 The population interventional indirect effect, $\E(Y_{1,G_1})$-$\E(Y_{1,G_0})$, compares the expected average counterfactual outcomes under hypothetical interventions in which the treatment is fixed but the mediator is changed from $G_1$ to $G_0$. This corresponds to the effect of treatment on the outcome operating through $M$, and is discussed further in \citet{vanderweele2017mediation}. 
  In terms of our MTO research question, the population interventional indirect effect is interpreted as the average difference in predicted risk of adolescent alcohol use setting voucher to be received and drawing mediator values from their counterfactual joint distribution if the voucher had been received versus had it not been received; or more simply, the effect of voucher receipt on the long-term risk of adolescent alcohol use that operates through aspects of the school, neighborhood, and social environments.
 
The subgroup predicted to \textit{not} have a harmful indirect effect (where ``harmful'' is indicated by a positive effect, as described above) can be identified by the target parameter \[\d(v) = \one(\B(v) \le 0),\] where $\B(v)$ is the average indirect effect conditional on a subset of baseline variables $V\subseteq W$, and is defined as
\[\B(v) = \E(Y_{1,G_1} - Y_{1,G_0}\mid V=v).\]
In what follows we let $W^c = W\backslash V$. The function $\B$ is often referred to as a ``blip'' function \citep{robins2004proceedings}, and  is a predictive function that takes covariates $V$ as input and outputs the conditional indirect effect of treatment on an additive scale. The function $\d(v)$ is then an indicator of a predicted nonharmful indirect effect for covariate profile $V=v$. In this case, if $\B(v) \le 0,$ then individuals in  strata $V=v$ are indicated as \textit{not} having a predicted harmful effect; if $\B(v) > 0,$ then individuals are indicated as having a predicted harmful effect. 

We introduce the following assumptions, which will allow us to identify the causal parameters $\B(v)$ and $\d(v)$:
\begin{assumption}[No unmeasured confounders of the $A\rightarrow Y$ relation]
$Y_{1,m} \indep A\mid W$\label{ass:AY}
\end{assumption}
\begin{assumption}[No unmeasured confounders of the $A\rightarrow M$ relation]
$M_{a} \indep A\mid W$, for $a \in \{0,1\}$.\label{ass:AM}
\end{assumption}
\begin{assumption}[No unmeasured confounders of the $M\rightarrow Y$ relation]
$Y_{1,m} \indep M\mid (W,A,Z)$.\label{ass:MY}
\end{assumption}
\begin{assumption}[Positivity]
$\p(W)>0$ implies $\p(a \mid W)>0$, $\p(M \mid  a^\star,W)>0$ and $\p(Z \mid  a',W)>0$ imply $\p(M \mid  a',Z,W)>0$ with probability one for $(a', a^\star)=(1,1)$ and $(a', a^\star)=(1,0)$. \label{ass:pos}
\end{assumption}

In the context of the conditional indirect effect of MTO housing voucher receipt on subsequent risk of adolescent alcohol use through the school, neighborhood and social environments, Assumptions 1 and 2 are expected to hold, as the intervention is randomized. Assumption 3 may not hold, most likely due to unmeasured post-randomization confounding variables, such as those related to changes in parental employment, income, parent-child dynamics, etc. However, we do include a large number of baseline covariates related parental socioeconomic status, motivations for enrolling in MTO, and relationships; child characteristics; and neighborhood characteristics. We also include an indicator representing moving with the housing voucher, which could be one such post-treatment confounder. Lastly, we include numerous mediating variables capturing aspects of the school, neighborhood and social environments. Any unmeasured variables would need to contribute to mediator-outcome confounding independently of the aforementioned variables to violate Assumption 3. 

Under Assumptions~\ref{ass:AY}-\ref{ass:pos}, the conditional indirect effect is identified from the observed data as follows. First, \citet{vanderweele2017mediation} show that for any values $(a',a^*)\in\{0,1\}^2$, $\E(Y_{a',G_{a^*}}\mid V=v)$ is identified as
\begin{equation*}
\E(Y_{a',G_{a^*}}\mid V=v) = \int \bb(a',z,m,w)\q(z\mid
  a',w)\p(m\mid a^*,w)\p(w^c\mid v)\dd \nu(w^c,z,m).
\end{equation*}
And thus we have
\begin{multline} \B(v) = \int \bb(1,z,m,w)\q(z\mid
  1,w)\{\p(m\mid 1,w)-\p(m\mid 0,w)\}\p(w^c\mid v)\dd \nu(w^c,z,m).\label{eq:idenB} 
\end{multline} 

\paragraph{Alternative interpretation in terms of the optimal dynamic treatment rule.} 
Although in the context of identifying boys in MTO who are predicted to have harmful indirect effects, we believe it would be inappropriate to interpret $\d(v)$ as a treatment rule, we recognize that other contexts exist in which such an interpretation may be justified. In these cases, $\d(v)$ may be alternatively defined in terms of an optimal individualized dynamic treatment rule: 
\begin{equation*}
    \d(v) \in \argmin_{\d' \in \mathcal D} \E(Y_{\d',G_{\d'}}) - \E(Y_{\d',G_0}),
\end{equation*}
where, for any $\d'$ and $\d^\star$, $Y_{\d',G_{\d^\star}}$ is the counterfactual outcome in a hypothetical world where treatment is set to the rule $\d'(V)\in\{0,1\}$ and the mediator is set to a random draw $G_{\d^\star}$ from the distribution of $M$ conditional on $A=\d^\star(V)$. Here $\mathcal D$ is the space of all functions that map the covariates $V$ to a treatment decision rule in $\{0,1\}$. 

\subsection{Estimation}
\label{sec:ruleest}
As pointed out in the Introduction, estimation of $\B(v)$ and $\d(v)$ are equivalent to estimation of conditional average effects (here, the conditional interventional indirect effect) and estimation of optimal treatment rules. Consequently, estimation techniques such as Q-learning \citep{Murphy03,qian2011performance,moodie2012q,laber2014interactive}, outcome-weighted learning (OWL) \citep{zhang2012estimating,zhao2012estimating,zhao2015new}, and doubly robust techniques \citep{luedtke2016super,diaz2018targeted,kennedy2020optimal} may be used. Q-learning is a regression-based approach related to g-computation that relies on sequential regression formulas. Q-learning is not directly applicable to our problem because Equation~\ref{eq:idenB} does not readily yield a sequential regression representation. OWL uses inverse probability weights to recast the problem of estimating $\d(v)$ as a weighted classification problem. Although we could use OWL to estimate $\B(v)$ and $\d(v)$ here, we instead choose to use a doubly robust approach that combines regressions with inverse probability weights to obtain an estimator that remains consistent under certain configurations of inconsistent estimation of nuisance parameters. This approach relies on so-called \textit{unbiased transformations}, which we define below. 

\begin{definition}[Unbiased transformation]\label{def:unbd}
The function $D(o)$ is an unbiased transformation for $\B(v)$ if it satisfies $\E[D(O)\mid V=v] = \B(v)$.
\end{definition}

In particular, we use a multiply robust unbiased transformation \citep{rubin2007doubly}. For the case of optimal treatment rules, multiply robust unbiased transformations are constructed using the efficient influence function (EIF) of the average treatment effect \citep{luedtke2016super,diaz2018targeted,kennedy2020optimal}. We take a similar approach here. Specifically, the EIF for the counterfactual mean $\E(Y_{a', G_a^*})$ under Assumptions~\ref{ass:AY}-\ref{ass:pos} was derived by \citet{diaz2019non} as follows. Define
   \begin{align*}
    \h(z, m, w)&=\frac{\g(a'\mid w)}{\g(a^{\star}\mid w)}
                    \frac{\q(z\mid a',w)}{\rr(z\mid a',m,w)}
      \frac{\e(a^{\star}\mid m, w)}{\e(a'\mid m,w)}\\\
    \uu(z,w) &= \E\left\{\bb(a',Z,M,W)\h(Z,M,W),\mid \,
      Z=z,A=a',W=w\right\},\\
    \vv(w) &= \E\left\{\int_{\mathcal
               Z}\bb(a',z,M,W)\q(z\mid a',W)\dd\nu(z)\,\mid \, A=a^{\star},W=w\right\}.
  \end{align*}
Let $\eta=(\g, \e, \q, \rr, \bb, \uu, \vv)$ denote a vector of nuisance parameters. Then the EIF for $\E(Y_{a', G_a^*})$ under Assumptions~\ref{ass:AY}-\ref{ass:pos} is equal to
\begin{align*}
  D_\eta^{(a^\prime,a^\star)}(o) &= \frac{\one\{a=a'\}}
  {\g(a'\mid w)}\h(z,m,w)\{y - \bb(a',z,m,w)\}\\
                        & + \frac{\one\{a=a'\}}{\g(a'\mid w)}\{\uu(1,w)-
                        \uu(0,w)\}\left\{z -
                          \q(1\mid a',w)\right\}\\
                        & + \frac{\one\{a=a^{\star}\}}{\g(a^{\star}
                        \mid w)}\left\{\int\bb(a',z,m,w)\q(z
                        \mid a',w)\dd\nu(z)-\vv(w)\right\},\\
            & + \vv(w).
\end{align*}

Thus, we use $D_{\eta}(o) = D_{\eta}^{(1,1)}(o) - D_{\eta}^{(1,0)}(o)$ as an unbiased transformation for $\B(v)$. For a fixed $\eta_1=(\g_1, \e_1, \q_1, \rr_1, \bb_1, \uu_1, \vv_1)$, \citet{diaz2019non} show that $D_{\eta_1}(o)$ satisfies Definition~\ref{def:unbd} as long as 
  \begin{enumerate}
  \item $\vv_1=\vv$ and either $(\q_1,\e_1,\rr_1)=(\q,\e,\rr)$
    or $(\bb_1,\q_1)=(\bb,\q)$,  or
  \item $\g_1=\g$ and either $(\q_1,\e_1,\rr_1)=(\q,\e,\rr)$
    or $(\bb_1,\q_1)=(\bb,\q).$
  \end{enumerate}
In this sense, using a multiply robust unbiased transformation provides robustness to misspecification of some models, hence the name. 

Our estimator proceeds by obtaining an estimate, $\hat\eta$, of $\eta$. All nuisance parameters are estimated using an ensemble regression approach known as the Super Learner \citep{van2007super,vanderLaanDudoitvanderVaart06}, which creates an optimally weighted combination of algorithms.  
 Then, the pseudo-outcome $D_{\hat\eta}(O_i)$ is computed for all individuals in the sample, and a regression of the pseudo-outcome on baseline covariates $V$ is fitted. Following the approach of \citet{luedtke2016super}, we also use Super Learner in fitting the regression of $D_{\hat\eta}(O)$ on $V$ and use the fitted values to obtain predictions  
 $\hat \B(v)$. The group of individuals at risk for a harmful indirect effect is identified as those indices $i$ with $\hat \d(v_i)=\one\{\hat \B(v_i)>0\}$.

We use cross-fitting \citep{klaassen1987consistent,zheng2011cross, chernozhukov2016double} throughout the estimation process in all fits. Let ${\cal V}_1, \ldots, {\cal V}_J$
denote a random partition of data with indices $i \in \{1, \ldots, n\}$ into $J$
prediction sets of approximately the same size such that 
$\bigcup_{j=1}^J {\cal V}_j = \{1, \ldots, n\}$. For each $j$,
the training sample is given by
${\cal T}_j = \{1, \ldots, n\} \setminus {\cal V}_j$. 
$\hat \eta_{j}$ denotes the estimator of $\eta$, obtained by training
the corresponding prediction algorithm using only data in the sample
${\cal T}_j$, and $j(i)$ denotes the index of the
validation set which contains observation $i$. We then use these fits, $\hat\eta_{j(i)}(O_i)$ in computing $D_{\hat\eta_{j(i)}}(O_i)$. Likewise, regressions of the pseudo-outcome $D_{\hat\eta_{j(i)}}(O_i)$ on $V_i$ are trained within each training sample, and the final estimate is computed by predicting in the corresponding validation data set. 

\section{Estimating the indirect effect under a hypothetical treatment decision $\d(v)$}
\label{sec:estim}
As stated in the Introduction, our goal is to identify a subset of the population that is predicted to have a harmful indirect effect, and for whom treatment may be carefully considered or reconsidered. Even though our goal is not to develop a treatment rule, in some situations it may be important to assess the population effects that would be observed if the function $\d$ were used as a treatment rule. To that end, in this section we estimate the hypothetical population interventional indirect effect if we were to use $\d(V)$ to assign treatment to each individual. Specifically, we define the total effect of implementing $\d(V)$ as $\E(Y_\d - Y_0)$, and decompose it in terms of direct and indirect effects as

\[\E(Y_\d - Y_0) = \underbrace{\E(Y_{\d,G_{\d}} - Y_{\d,G_0})}_{\text{indirect effect}} + \underbrace{\E(Y_{\d,G_0} - Y_{0,G_{0}})}_{\text{direct effect}}.\]
The population interventional indirect effect can be identified under the sequential randomization assumptions and positivity in Section \ref{sec:ruleident} as: 
\begin{multline} \E(Y_{\d, G_{\d}} - Y_{\d, G_{0}}) =\\ \int \bb(\d(v),z,m,w)\q(z\mid
  \d(v),w)\{\p(m\mid \d(v),w)-\p(m\mid 0,w)\}\p(w)\dd \nu(w,z,m).\label{eq:identeval} 
\end{multline} 
Because we do not have the true, unknown $\d$, but instead have an estimate of it, $\hat\d$, we estimate the population indirect effect $\E(Y_{\hat{\d},G_{\hat{\d}}} - Y_{\hat{\d},G_0})$ that would be observed if our estimate were implemented. We use the one-step estimation approach described in \citet{diaz2019non}, and which is based on solving the EIF estimating equation. Specifically, we let our estimator be defined as
    \[\hat\theta=\frac{1}{n}\sum_{i=1}^n \{D_{\hat{\eta}}^{\hat{\d},\hat{\d}}(O_i) - D_{\hat{\eta}}^{\hat{\d},0}(O_i)\}, \] 
 We use a cross-fitted version of this estimator, as described in Section \ref{sec:ruleest}. The variance of this estimator can be estimated as the sample variance of the EIF. Theorem 5 of \citet{van2015targeted} proves that plugging in $\hat\d$ estimated from the data results in asymptotically linear estimation of its effect, 
 even though the same data were used to estimate $\hat{\d}$ and to assess its effect. 
 

The R code to implement this cross-fitted estimator is available:\\
\url{blinded for review}. 

\section{Identifying those with predicted harmful indirect effects}
We now apply our proposed approach to identify the subgroup of boys with a predicted harmful indirect effect of voucher receipt on adolescent alcohol use through aspects of the school, neighborhood, and social environments. In this case, the average treatment effect of Section 8 housing voucher receipt on long-term risk of adolescent alcohol use among boys is beneficial, contributing to \textit{decreased} risk, whereas the indirect effect through mediators related to the school and neighborhood environments and instability of the social environment is estimated to contribute to an \textit{increased} risk of subsequent adolescent alcohol use. There is evidence that improvements in measures of the school and neighborhood environment (e.g., neighborhood poverty, school rank, school poverty) may actually negatively impact a subgroup of boys, possibly via increased risk of suspensions/expulsions and less social support \citep{rudolph2020helped}, which in turn, may increase risk of a variety of negative mental health and substance use outcomes \citep{rudolph2020helped}.

If one is interested in optimizing the overall effect, one could apply existing methods to identify the subgroup who may be harmed by the intervention \citep{luedtke2016super,robins2004proceedings,zhao2012estimating}. However, instances where the overall effect is beneficial, but the indirect effect is harmful---like this  one---motivate our proposed approach to identify a subgroup predicted to experience an unintended, harmful indirect effect, even in the presence of a predicted beneficial overall effect. In this application, we 1) identify the subgroup of boys who would be predicted to have a harmful indirect effect and thus, for whom voucher receipt may not be recommended, and 2) estimate the hypothetical indirect effect and hypothetical total effect had the identified subgroup not received the intervention. 

A limitation of using so-called ``black-box'' algorithms proposed in Section \ref{sec:ruleest} for this task is that we are left without knowing which characteristics are important in predicting whether an indirect effect will be positive or negative. Consequently, we also use a 
 recently proposed adaptive lasso \citep{bahamyirou2020doubly} to learn less-than-optimal, but interpretable $\d$. 

\subsection{Data and Analysis}
MTO was a large-scale experiment in which families living in low-income public housing could sign-up to be randomized to receive a Section 8 housing voucher, which is a form of housing assistance that subsidizes rent on the private market, allowing families to move out of public housing. We restrict to adolescent boys who were surveyed at the final follow-up timepoint in the Boston, Chicago, New York City, and Los Angeles sites (N=2,100, rounded sample size). We consider baseline covariates at the individual, family, and neighborhood level ($W$, study site, age, race/ethnicity, number of family members, previous problems in school, enrolled in special class for gifted and talented students, parent is high school graduate, parent marital status, parent work status, receipt of AFDC/TANF, whether any family member has a disability, perceived neighborhood safety, neighborhood satisfaction, neighborhood poverty level, reported reasons for participating in MTO, previous number of moves, previous application for Section 8 voucher), baseline randomized intervention status of received voucher or not ($A$, binary 1/0), whether or not the family used the voucher to move ($Z$, binary 1/0), mediating variables representing aspects of the school and neighborhood environments and instability of the social environment over the 10-15 years of follow-up ($M$, school rank, student-to-teacher ratio, \% students receiving free or reduced-price lunch, \% schools attended that were Title I, number of schools attended, number of school changes within the year, whether or not the most recent school was in the same district as the baseline school, number of moves, neighborhood poverty; all weighted over the duration of follow-up), and the long-term outcome of past-month alcohol use at the final timepoint, when the children were adolescents ($Y$, binary 1/0). For the purposes of this illustration, we use just one imputed dataset. As recommended for all MTO analyses, we use individual-level weights that account for the sampling of children within families, assignment ratios, and loss-to-follow-up \citep{sanbonmatsu2011moving}.

We first apply the estimation approach in Section \ref{sec:ruleest} to identify the subgroup $\{i\in\{1,\ldots,n\}:\hat\d(V_i)=0\}$ of boys with a predicted harmful indirect effect of voucher receipt on adolescent alcohol use through aspects of the school, neighborhood, and social environments.  We then apply the estimation procedure in Section \ref{sec:estim} to estimate indirect effects: $\E(Y_{\hat\d,G_{\hat\d}}-Y_{\hat\d,G_{0}})$, $\E(Y_{1,G_{1}} - Y_{1,G_{0}})$
, and total effects: $\E(Y_{\hat\d,G_{\hat\d}} - Y_{0,G_{0}})$, $\E(Y_{1,G_{1}} - Y_{0,G_{0}})$. 
We use five folds in cross-fitting, and include: a simple mean model, main-terms generalized linear model, lasso, multivariate adaptive regression splines, and extreme gradient boosted machines in our Super Learner ensemble. Finally, we report the estimated indirect effects and total effects using the alternative, interpretable $\hat{\d}$ estimated using 
 adaptive lasso.

\subsection{Results}

\begin{figure}[h!]
\caption{Population interventional indirect effects and total effects by rule type. All results were approved for release by the U.S. Census Bureau, authorization number CBDRB-FY21-ERD003-004.}
\label{fig:res}
\centering
  \includegraphics[width=0.75\textwidth]{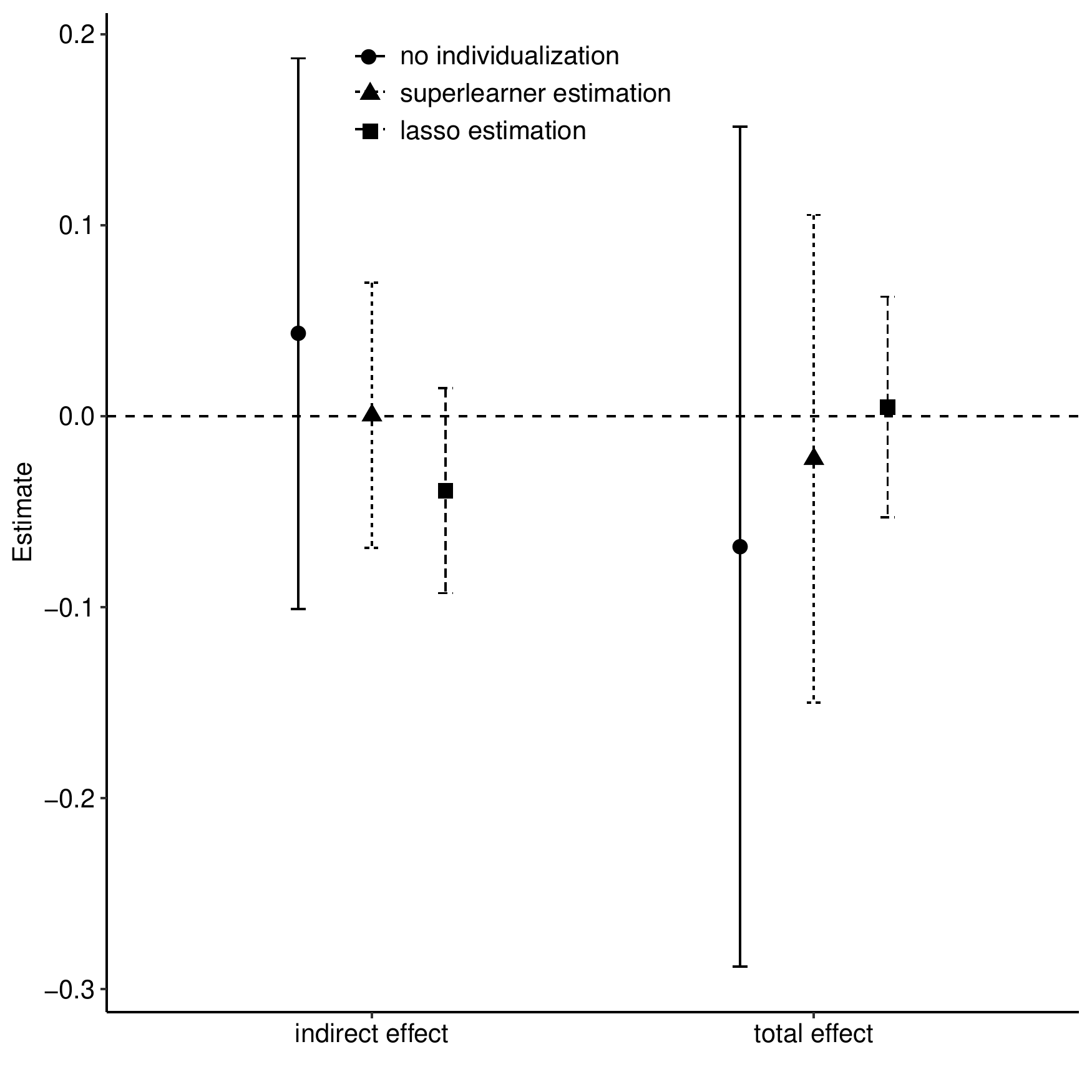}

\end{figure}

Figure \ref{fig:res} shows: the typical population interventional indirect effect (PIIE, $\E(Y_{1,G_{1}} - Y_{1,G_{0}})$) estimates and total effect (PITE, $\E(Y_{1,G_{1}} - Y_{0,G_{0}})$) estimates not using any individualization, which are labeled as ``no individualization'', and the PIIE and PITE estimates using $\hat\d$ to hypothetically assign boys to receive the voucher who were not predicted to have a harmful experience through mediators of the school, neighborhood, and social environments (learned via Super Learner), denoted ``superlearner estimation''. Using Super Learner to identify the subgroup predicted to experience a harmful indirect effect and then not giving the voucher to that subgroup would be expected to reduce the otherwise harmful indirect effect down to 0.0004 increased risk of subsequent adolescent alcohol use (95\% CI: -0.0690, 0.0698) as opposed to 0.0432 increased risk (95\% CI: -0.1009, 0.1873). It would also be expected to slightly decrease the beneficial total average effect of voucher assignment to 0.0224 reduced risk of subsequent adolescent alcohol use (-0.0224, 95\% CI: -0.1500, 0.1053) as opposed to 0.0683 reduced risk (95\% CI: -0.2882, 0.1053). 

 Applying the adaptive lasso to estimate $\hat{d}$, we find that those whose parent graduated from high school and whose race is not white, black, nor Hispanic/Latino would be predicted to not have a harmful indirect effect. The PIIE and PITE estimates using this $\hat{\d}$ are denoted 
 ``lasso estimation''. 

\section{Conclusions}
We proposed a nonparametric approach to identify a subgroup who may have unintended harmful indirect effects even in the presence of beneficial overall effects. We then proposed a nonparametric estimator to estimate the hypothetical population interventional indirect and total effects were one to make the decision not to treat the subgroup predicted to have harmful indirect effects. Our cross-fitted estimators solve the efficient influence function, so are robust and can incorporate machine learning in model fitting. 

This work was motivated by surprising results from a large, housing policy intervention (the Moving to Opportunity Study) in which boys whose families were randomized to receive a housing voucher had slightly reduced risk of subsequent alcohol use as adolescents (total effect) but the indirect pathway from voucher receipt to that positive effect operating through aspects of the school and social environments was harmful. This harmful indirect effect could reflect processes like boys whose families moved with the vouchers going to schools with higher academic performance but where they were more likely to face discriminatory practices, like suspension and expulsion, and more social isolation \citep{rudolph2020helped}. It is possible that some may wish to avoid such negative processes, prompting the desire to identify those at risk for them. In this example, we do so and find that not giving the intervention to this subgroup would have indeed reduced the harmful predicted indirect effect point estimate, though the confidence intervals are wide and overlapping. 

In terms of identification and estimation, our proposal builds on both the causal mediation and individualized optimal treatment regime literatures. However, we caution against interpreting $\hat{\d}$ as a decision rule, understanding that, at least in some instances,  
 deciding whether or not someone would benefit from an intervention/treatment is necessarily 
 more nuanced than the mediation mechanisms and total effects we consider. 
 Future work could focus on incorporating additional 
 complexities and nuances into the existing tools used in personalized medicine to reflect the multiple mechanisms by which treatments or interventions may affect numerous, and sometimes competing, outcomes among individuals, recognizing heterogeneities (anticipated or not) at each stage in the process. 









\bibliographystyle{Chicago}

\bibliography{lib}
\end{document}